\documentclass[%
 reprint,
 amsmath,amssymb,
 aps,
 prl
]{revtex4-1}

\usepackage{graphicx}
\usepackage{dcolumn}
\usepackage{bm}
\usepackage{color}
\usepackage{tikz}
\usetikzlibrary{matrix}
\usepackage{tabularx}
\usepackage{arydshln,leftidx,mathtools}
\usepackage{mathdots}

\begin{document}

\preprint{APS/123-QED}

\title{Anomalous Proximity Effect of Planer Topological Josephson Junctions}

\author{S. Ikegaya$^{1}$, S. Tamura$^{2}$, D. Manske$^{1}$, and Y. Tanaka$^{2}$}
\affiliation{$^{1}$Max-Planck-Institut f\"ur Festk\"orperforschung, Heisenbergstrasse 1, D-70569 Stuttgart, Germany\\
$^{2}$Department of Applied Physics, Nagoya University, Nagoya 464-8603, Japan}

\date{\today}

\begin{abstract}
The anomalous proximity effect in dirty superconducting junctions
is one of most striking phenomena highlighting the profound nature of Majorana bound states and odd-frequency Cooper pairs in topological superconductors.
Motivated by the recent experimental realization of planar topological Josephson junctions,
we describe the anomalous proximity effect in a superconductor/semiconductor hybrid,
where an additional dirty normal-metal segment is extended from a topological Josephson junction.
The topological phase transition in the topological Josephson junction is accompanied by a drastic change in the low-energy transport properties of the attached dirty normal-metal.
The quantization of the zero-bias differential conductance, which appears only in the topologically nontrivial phase,
is caused by the penetration of the Majorana bound states and odd-frequency Cooper pairs into a dirty normal-metal segment.
As a consequence, we propose a practical experiment for observing the anomalous proximity effect.
\end{abstract}

\maketitle

Majorana bound states (MBSs) in topological superconductors~\cite{kane_10,zhang_11,tanaka_12,sato_17},
which have opened a promising avenue for the realization of fault-tolerant quantum computations~\cite{ivanov_01,sarma_08, sau_11},
have recently become a focus of intense research in condensed matter physics.
For the past decade, the existence of MBSs has been experimentally demonstrated in various topologically nontrivial superconducting systems including
semiconductor/superconductor hybrids
~\cite{sarma_10,oreg_10,kouwenhoven_12,deng_12,kouwenhoven_18,halperin_17,haim_19,setiawan_19,nichele_19,yacoby_19,shabani_20,marcus_20},
magnetic atom chains fabricated on superconductors~\cite{beenakker_11,yazdani_13,yazdani_14},
and superconducting topological insulators~\cite{kane_08,tanaka_09,tanaka_10,cava_10,ando_11,zhang_10,wang_17}.
Based on this significant progress, the time has come to go beyond proving the existence of MBSs and investigate their deep characteristics more thoroughly.

One of the most striking phenomena caused by MBSs is an anomalous proximity effect in dirty superconducting junctions.
When a dirty normal-metal (DN) is attached to a topological superconductor, the MBS penetrates into the attached DN and induces various anomalies
including the formation of a zero-energy peak in the local density of states of the attached DN~\cite{tanaka_04,tanaka_05(1),tanaka_05(2),tanaka_07,higashitani_09},
the zero-bias conductance quantization in a DN/superconductor junction~\cite{tanaka_04,asano_07,ikegaya_15,ikegaya_16(1)},
and the fractional Josephson effect in a superconductor/DN/superconductor junction~\cite{asano_06,ikegaya_16(2)}.
Moreover, it has been shown that there is an essential duality between the MBSs and odd-frequency Cooper pairs~\cite{asano_13,tamura_19},
where pair functions of odd-frequency Cooper pairs have an odd parity with respect to time (frequency)~\cite{tanaka_12,berezinskii_74,linder_19,cayao_20}.
Thus, the penetration of MBSs into the DN simultaneously means that odd-frequency Cooper pairs are formed in the attached DN~\cite{tanaka_07,asano_13}.
Although the first theoretical prediction for the anomalous proximity effect was made over 15 years ago~\cite{tanaka_04},
this effect has not been observed experimentally owing to a lack of candidate materials hosting the MBSs.
Nevertheless, the recent and rapid progress achieved in the fabrication techniques used in topological superconductors have shed some light on this issue.

\begin{figure}[bbbb]
\begin{center}
\includegraphics[width=0.42\textwidth]{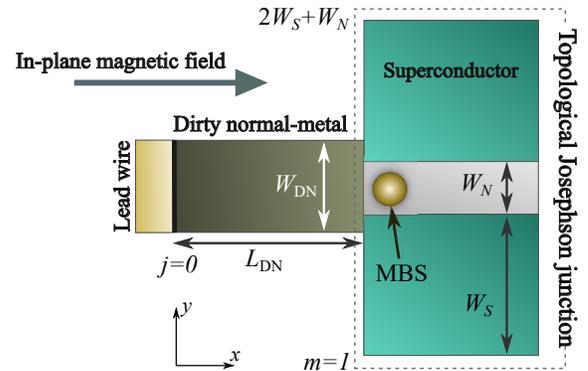}
\caption{Schematic image of the planar topological Josephson junction with the additional dirty normal-metal segment.}
\label{fig:figure1}
\end{center}
\end{figure}

In this study, we focus on a planer topological Josephson junction (TJJ)~\cite{halperin_17,haim_19,setiawan_19}
realized in recent experiments~\cite{nichele_19,yacoby_19,shabani_20}.
The MBSs of a TJJ originate not from the band topology of the \textit{bulk states} as is typically the case,
but from the non-trivial band topology of the \textit{Andreev bound states} appearing within the vicinity of the junction interface.
Owing to this peculiarity, it remains unclear whether a TJJ has an anomalous proximity effect.
To solve this ambiguity, we studied the differential conductance of a TJJ with an additional DN segment, as shown in Fig.~\ref{fig:figure1}.
Herein, we demonstrate that the minimum value of the zero-bias differential conductance is quantized to $2e^2/h$ only during a topologically nontrivial phase (Fig.~\ref{fig:figure3}).
In addition, we discuss the penetration of the MBSs (Fig.~\ref{fig:figure4}) and odd-frequency $s$-wave Cooper pairs (Fig.~\ref{fig:figure5}) into the attached DN.

Notably, the proposed system is fabricated on a thin film semiconductor, the microfabrication techniques of which are well established.
Moreover, the conductance spectrum changes drastically through a topological phase transition, which is driven simply by changing the superconducting phase difference.
As a consequence, we propose a practical experiment for observing the anomalous proximity effect,
which is a crucial subject in the physics of both MBSs and odd-frequency Cooper pairs.

\textit{Model.}
In this study, we consider a semiconductor/superconductor hybrid, as shown in Fig.~\ref{fig:figure1}.
Two spin-singlet $s$-wave superconductors are fabricated on a thin film semiconductor allowing the construction of a planer TJJ.
Moreover, an additional normal-metal segment is extended from the TJJ segment.
The extended normal-metal segment contains disordered potentials that can be introduced for instance by a focused ion beam technique~\cite{muroe_04,stevie_05}.
Herein, we describe the present system using a tight-binding model, where a lattice site is indicated by a vector $\boldsymbol{r}=j \boldsymbol{x} + m \boldsymbol{y}$.
The junction consists of three segments:
a lead wire ($-\infty \leq j \leq 0$), a DN segment ($1 \leq j \leq L_{\mathrm{DN}}$), and a TJJ segment ($L_{\mathrm{DN}}+1 \leq j \leq \infty$).
In the $y$ direction, the TJJ is located between $1 \leq m \leq 2W_S + W_N$ with $W_{S(N)}$ representing the width of the superconducting (normal) region.
The width of the DN segment and lead wire is given by $W_{\mathrm{DN}}$, where the center of the DN segment is aligned with that of the TJJ segment.
The TJJ is described using a Bogoliubov--de Gennes (BdG) Hamiltonian, $H=H_N + H_{\Delta}$, with
\begin{align}
H_N =& -t \sum_{\langle \boldsymbol{r}, \boldsymbol{r}^{\prime} \rangle,\sigma}
\left[c^{\dagger}_{\boldsymbol{r},\sigma}c_{\boldsymbol{r}^{\prime},\sigma} + \mathrm{h.c.} \right] 
-\mu \sum_{\boldsymbol{r},\sigma}c^{\dagger}_{\boldsymbol{r},\sigma}c_{\boldsymbol{r},\sigma} \nonumber\\
&+\frac{i \lambda}{2}\sum_{\boldsymbol{r},\sigma,\sigma^{\prime}}\left( \sigma_y \right)_{\sigma,\sigma^{\prime}}
\left[c^{\dagger}_{\boldsymbol{r}+\boldsymbol{x},\sigma}c_{\boldsymbol{r},\sigma^{\prime}}
-c^{\dagger}_{\boldsymbol{r},\sigma} c_{\boldsymbol{r}+\boldsymbol{x},\sigma^{\prime}} \right] \nonumber\\
&-\frac{i \lambda}{2}\sum_{\boldsymbol{r},\sigma,\sigma^{\prime}}\left( \sigma_x \right)_{\sigma,\sigma^{\prime}}
\left[c^{\dagger}_{\boldsymbol{r}+\boldsymbol{y},\sigma}c_{\boldsymbol{r},\sigma^{\prime}}
-c^{\dagger}_{\boldsymbol{r},\sigma} c_{\boldsymbol{r}+\boldsymbol{y},\sigma^{\prime}} \right] \nonumber\\
&+V_Z \sum_{\boldsymbol{r},\sigma,\sigma^{\prime}} \left( \sigma_x \right)_{\sigma,\sigma^{\prime}}
c^{\dagger}_{\boldsymbol{r},\sigma}c_{\boldsymbol{r},\sigma^{\prime}}, \\
H_{\Delta}=&\sum_j \sum_{m=1}^{W_S} \left[\Delta e^{i \varphi/2}
c^{\dagger}_{\boldsymbol{r},\uparrow}c^{\dagger}_{\boldsymbol{r},\downarrow} + \mathrm{h.c.}\right] \nonumber\\
&+\sum_j \sum_{m=1+W_S}^{W_N+W_S} \left[\Delta e^{-i \varphi/2} c^{\dagger}_{\boldsymbol{r},\uparrow}c^{\dagger}_{\boldsymbol{r},\downarrow} + \mathrm{h.c.}\right],
\end{align}
where $c^{\dagger}_{\boldsymbol{r},\sigma}$ ($c_{\boldsymbol{r},\sigma}$) is the creation (annihilation) operator of an electron
at $\boldsymbol{r}$ with spin $\sigma$ ($=\uparrow$, $\downarrow$),
$t$ denotes the nearest-neighbor hopping integral, and $\mu$ is the chemical potential.
The strength of the Rashba spin-orbit coupling is represented by $\lambda$.
The Zeeman potential induced by the externally applied magnetic field in the $x$ direction is given by $V_Z$.
The amplitude of the pair potential is denoted as $\Delta$, where $\varphi$ represents the superconducting phase difference between the two superconducting segments.
The Pauli matrices in the spin space are represented by $\sigma_{\nu}$ ($\nu=x$, $y$, and $z$).
The DN segment is described using $H_{\mathrm{DN}}=H_N + H_{\mathrm{DP}}$ with
\begin{align}
H_{\mathrm{DP}} = \sum_{\boldsymbol{r},\sigma}v(\boldsymbol{r}) c^{\dagger}_{\boldsymbol{r},\sigma}c_{\boldsymbol{r},\sigma},
\end{align}
where $v(\boldsymbol{r})$ is the disordered potential given randomly within the range of $-X \leq v(\boldsymbol{r}) \leq X$.
The lead wire is described as follows:
\begin{align}
H^{\prime} =&  -t^{\prime} \sum_{\langle \boldsymbol{r}, \boldsymbol{r}^{\prime} \rangle,\sigma}
\left[c^{\dagger}_{\boldsymbol{r},\sigma}c_{\boldsymbol{r}^{\prime},\sigma} + \mathrm{h.c.} \right] 
-\mu^{\prime} \sum_{\boldsymbol{r},\sigma}c^{\dagger}_{\boldsymbol{r},\sigma}c_{\boldsymbol{r},\sigma} \nonumber\\
&+V^{\prime}_Z \sum_{\boldsymbol{r},\sigma,\sigma^{\prime}} \left( \sigma_x \right)_{\sigma,\sigma^{\prime}}
c^{\dagger}_{\boldsymbol{r},\sigma}c_{\boldsymbol{r},\sigma^{\prime}},
\end{align}
where $t^{\prime}$, $\mu^{\prime}$, and $V^{\prime}_Z$ represent the hopping integral, chemical potential, and Zeeman potential in the lead wire, respectively.
We denote the hopping integral between the lead wire and the DN segment (i.e., the hopping integral between $j = 0$ and $j=1$) as $t_{\mathrm{int}}$.
A more detailed expression for the BdG Hamiltonian is given in Supplemental Material~\cite{supplemental}.
In the following calculations, we fix the parameters as $t = t^{\prime}=1.0$, $t_{\mathrm{int}}=0.1$, $\mu=-3.5$, $\mu^{\prime}=-3.0$,
$\lambda = 0.5$, $\Delta=0.1$, $X=1.0$, $W_S = 18$, $W_N=4$, $W_{\mathrm{DN}}=10$, and $L_{\mathrm{DN}}=20$.
In addition, we assume that the relation of $V_Z = V^{\prime}_Z$ holds.
For the random ensemble average, $10^4$ samples are used.

\begin{figure}[bbbb]
\begin{center}
\includegraphics[width=0.3\textwidth]{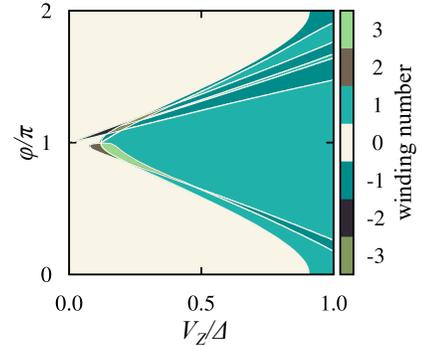}
\caption{Topological  phase diagram as a function of the Zeeman potential and superconducting phase difference.}
\label{fig:figure2}
\end{center}
\end{figure}
Before discussing the anomalous proximity effect, we briefly summarize the topological property of the TJJ~\cite{halperin_17}.
To evaluate the topological number, we remove the DN segment from the TJJ and apply a periodic boundary condition in the $x$ direction.
We represent the BdG Hamiltonian of the TJJ with momentum $k_x$ using $\check{H}(k_x)$,
where the explicit form of $\check{H}(k_x)$ is given in Supplemental Material~\cite{supplemental}.
The TJJ intrinsically has a particle--hole symmetry as $\check{C} \, \check{H}(k_x) \, \check{C}^{-1} = -\check{H}(-k_x)$ with $\check{C}^2=+1$.
In addition, the TJJ preserves the time-reversal symmetry as $\check{T}_+\, \check{H}(k_x) \, \check{T}_+^{-1} = \check{H}(-k_x)$,
where $\check{T}_+ = \check{M}_y \check{T}_-$ by satisfying $\check{T}_+^2=+1$.
Here, $\check{M}_y$ and $\check{T}_-$ represent a mirror symmetry operator with respect to the $x$-$z$ plane
and a conventional time-reversal symmetry operator satisfying $\check{T}_-^2=-1$, respectively.
Combining $\check{T}_+$ and $\check{C}_+$, the chiral symmetry of the TJJ is defined as follows:
$\check{S} \, \check{H}(k_x) \, \check{S} = - \check{H}(k_x)$, where $\check{S}=-\check{T}_+ \check{C}$.
The explicit forms for the symmetry operators are given in Supplemental Material~\cite{supplemental}.
Because $\check{T}_+^2=+1$ and $\check{C}^2=+1$, TJJ belongs to the BDI  symmetry class~\cite{schnyder_08}.
Thus, we can define a one-dimensional winding number by~\cite{sato_11,sau_12}
\begin{align}
w = \frac{1}{4 \pi i} \int dk_x \mathrm{Tr} \left[ \check{S} \check{H}^{-1}(k_x) \partial_{k_x} \check{H}(k_x) \right].
\end{align}
According to the bulk-boundary correspondence, we can expect the $|w|$ MBSs at the interface between the TJJ and DN (see Fig.~\ref{fig:figure1}).
Simultaneously, we can also define a $\mathbb{Z}_2$ topological number given by $\mathbb{Z}_2 = (-1)^{w}$~\cite{halperin_17}.
In fact, the mirror symmetry of $\check{M}_y$ is easily broken by perturbations such as impurities, and the symmetry class of the TJJ changes into class D.
However, even in the absence of mirror symmetry,
the TJJ with odd winding numbers can still exhibit a single MBS, which is actually characterized by the $\mathbb{Z}_2$ topological number.
Physically, the single MBS is protected by the particle--hole symmetry, which is preserved irrespective of the mirror symmetry.
In Fig.~\ref{fig:figure2},
we show the topological phase diagram with the present parameter choices as a function of the Zeeman potential and the superconducting phase difference.
We find the topologically nontrivial (topological) phases with various nonzero winding numbers, most of which belong to the odd winding numbers.

\textit{Anomalous proximity effect.}
We now consider the differential conductance in the present system, where electrons are injected from the lead wire.
Within the Blonder--Tinkham--Klapwijk (BTK) formalism, the differential conductance at zero temperature is calculated using~\cite{klapwijk_82, bruder_90, kashiwaya_00}
\begin{align}
G(eV) = \frac{e^{2}}{h} \sum_{\zeta,\zeta^{\prime}}
\left[ \delta_{\zeta,\zeta^{\prime}} - \left| r^{ee}_{\zeta,\zeta^{\prime}} \right|^{2}
+ \left| r^{he}_{\zeta,\zeta^{\prime}} \right|^{2} \right]_{E=eV},
\end{align}
where $r^{ee}_{\zeta,\zeta^{\prime}}$ and $r^{he}_{\zeta,\zeta^{\prime}}$ denote a normal and Andreev reflection coefficient at energy $E$, respectively.
The indexes $\zeta$ and $\zeta^{\prime}$ label an outgoing and incoming channel in the normal lead wire, respectively.
These reflection coefficients are calculated using lattice Green's function techniques~\cite{fisher_81, ando_91}.
We assume a sufficiently low transparency at the lead--wire/DN interface ($t_{\mathrm{int}}=0.1$) such that the bias voltage is mainly dropped at this interface
~\cite{tanaka_05(1), tanaka_05(2)}.
Based on this assumption, the BTK formalism is quantitatively justified for bias voltages well below the superconducting gap.
In Figs.~\ref{fig:figure3}(a) and \ref{fig:figure3}(b), we show the differential conductance for the topological phase with $w=+1$
and for the non-topological phase (i.e., $w=0$) as a function of the bias voltage, respectively
For the topological (non-topological) phase, we choose $V_Z = 0.6 \Delta$ ($0.2 \Delta$) and $\varphi = \pi$ ($0.1 \pi$).
As shown in Fig.~\ref{fig:figure3}(a), the conductance spectrum for the topological phase shows a zero-bias peak structure,
where the zero-bias differential conductance (ZBC) is almost $2e^2/h$~\cite{tanaka_04}.
By contrast, as shown in Fig.~\ref{fig:figure3}(b), the conductance spectrum for the non-topological phase shows an almost M-shaped structure,
where the conductance enhancement of approximately $eV=\pm 0.3\Delta$ is related to the Andreev bound states formed at the junction interface of the TJJ.
In Fig.~\ref{fig:figure3}(c), the ZBC is shown as a function of the Zeeman potential and superconducting phase difference.
We can see that the ZBC is almost $2e^2/h$ for the entire topological phase with the odd winding numbers,
whereas the ZBC for the non-topological phase and that for the topological phase with the even winding numbers are almost zero.
Strictly speaking, the ZBC in the topological phase with odd winding numbers is slightly greater than $2e^2/h$ because
the normal propagating channels, which do not couple with the MBS, can also contribute to the charge current.
Nevertheless, when the transparency from the lead wire to the TJJ is sufficiently low,
the contribution from a resonant transmission channel related with the MBS~\cite{beenakker_11(1),beenakker_11(2),beenakker_11(3)} becomes dominant. 
Therefore, the minimal value of the ZBC in the present junction is exactly quantized to $2e^2/h$~\cite{ikegaya_16(1)}.
The minimal conductance quantization disappears with the even winding numbers (i.e., $w =\pm 2$).
This implies that the two MBSs can no longer retain their degeneracy at zero-energy
because of the broken mirror symmetry owing to the disordered potentials in the DN segment.
\begin{figure}[tttt]
\begin{center}
\includegraphics[width=0.42\textwidth]{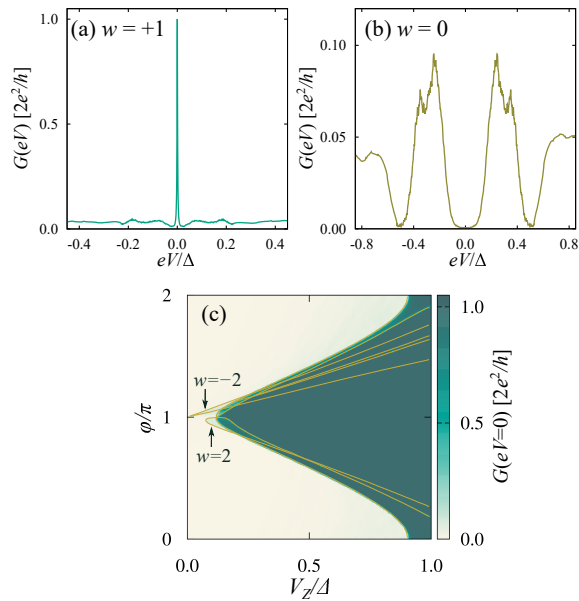}
\caption{Differential conductance for (a) the topological phase with $w=1$ and (b) non-topological phase as a function of the bias voltage.
(c) Zero-bias differential conductance as a function of the Zeeman potential and superconducting phase difference.
The yellow solid line denotes the topological phase boundary .}
\label{fig:figure3}
\end{center}
\end{figure}

Next, we discuss the local density of states (LDOS) in the DN segment.
The LDOS is calculated by the formula
$\rho(\boldsymbol{r},E) = - \mathrm{Tr}\left[ \mathrm{Im} \left\{ \check{G}(\boldsymbol{r},\boldsymbol{r},E+i\delta) \right\} \right]/\pi$,
where $\check{G}(\boldsymbol{r},\boldsymbol{r}^{\prime},E+i\delta)$ represents Green's function.
In addition, $\mathrm{Tr}$ indicates the trace in the spin and Nambu spaces; $\delta$ is a small imaginary part added to energy $E$.
In the following calculations, we fix $\delta=10^{-5}\Delta$.
In Fig.~\ref{fig:figure4}(a), we show the LDOS at zero energy as a function of the Zeeman potential and superconducting phase difference.
The LDOS is averaged in terms of the lattice sites in the DN segment as
$\langle \rho(E) \rangle_{\mathrm{DN}} = \sum_{\boldsymbol{r} \in \text{DN}} \rho(\boldsymbol{r},E)/S_{\mathrm{DN}}$ with
$S_{\mathrm{DN}}=W_{\mathrm{DN}} \times L_{\mathrm{DN}}$.
We can see that the zero energy LDOS in the DN segment suddenly increases when the system intersects the phase boundary
from the non-topological phase to the topological phase.
In Fig.~\ref{fig:figure4}(b), we show the LDOS for $1 \leq j \leq L_{\mathrm{DN}}$ as a function of the energy,
where we consider the topological phase with $V_Z = 0.5 \Delta$ and $\varphi = \pi$ (i.e., $w=+1$).
Here, we average the LDOS with respect to the lattice sites in the $y$ direction as
$\langle \rho(x,E) \rangle_{\mathrm{DN}} = \sum_{y \in \text{DN}} \rho(\boldsymbol{r},E)/W_{\mathrm{DN}}$.
We can see that the LDOS has a steep zero-energy peak structure within the entire DN.
The zero-energy peak structure appearing only in the topological phase implies that the MBS of the TJJ penetrates into the attached DN.
The minimal conductance quantization shown in Fig.~\ref{fig:figure3} is caused by the resonant transmission channel formed by
the penetrated MBS~\cite{ikegaya_15,ikegaya_16(1)}.
\begin{figure}[tttt]
\begin{center}
\includegraphics[width=0.48\textwidth]{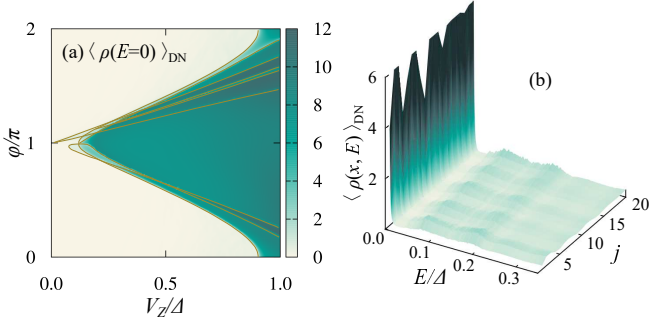}
\caption{(a) Local density of states at zero energy as a function of the Zeeman potential and superconducting phase difference.
(b) Local density of states in the topological phase with $w=+1$ as a function of the energy and position within the DN segment,
plotted within the range $\langle \rho(x,E) \rangle_{\mathrm{DN}}<6 $.}
\label{fig:figure4}
\end{center}
\end{figure}

\begin{figure}[bbbb]
\begin{center}
\includegraphics[width=0.48\textwidth]{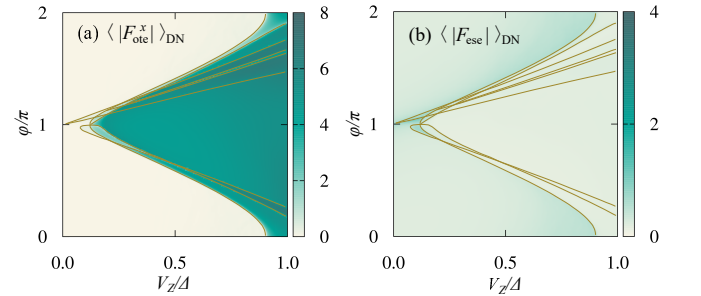}
\caption{Absolute value of (a) $F_{\mathrm{ote}}^{x}(\boldsymbol{r},\omega)$ and (b) $F_{\mathrm{ese}}(\boldsymbol{r},\omega)$
as a function of the Zeeman potential and superconducting phase difference.
The pair amplitudes are averaged in terms of the lattice sites in the DN segment,
where a Matsubara frequency of $\omega=10^{-4}\Delta$ is chosen.}
\label{fig:figure5}
\end{center}
\end{figure}
Finally, we discuss the odd-frequency Cooper pairs in the DN segment.
Here, we only focus on the pair amplitudes for the $s$-wave pairing symmetry because anisotropic pairings are intrinsically destroyed by the disordered potential.
According to the Fermi--Dirac statistics, there are two possible $s$-wave Cooper pairs.
The first pair has a conventional even-frequency spin-singlet even-parity (ESE) pairing symmetry whose pair amplitude is evaluated by the following:
\begin{align}
\hat{F}_{\mathrm{ese}}(\boldsymbol{r},\boldsymbol{r},\omega) 
&= \frac{\hat{F}(\boldsymbol{r},\boldsymbol{r},\omega) + \hat{F}(\boldsymbol{r},\boldsymbol{r},-\omega)}{2} \nonumber\\
&=\left[ \begin{array}{cc}
0 & F_{\mathrm{ese}}(\boldsymbol{r},\omega) \\
- F_{\mathrm{ese}}(\boldsymbol{r},\omega) & 0
\end{array}\right],
\end{align}
with $\hat{F}(\boldsymbol{r},\boldsymbol{r}^{\prime},\omega)$ being the anomalous part of the Matsubara Green's function with a Mastubara frequency $\omega$.
The other possible Cooper pair belongs to the odd-frequency spin-triplet even-parity (OTE) pairing symmetry, the pair amplitude of which is given by the following:
\begin{align}
\hat{F}_{\mathrm{ote}}(\boldsymbol{r},\boldsymbol{r},\omega) 
&= \frac{\hat{F}(\boldsymbol{r},\boldsymbol{r},\omega) - \hat{F}(\boldsymbol{r},\boldsymbol{r},-\omega)}{2}\nonumber\\
&=\left[ \begin{array}{cc}
-F_{\mathrm{ote}}^x + i F_{\mathrm{ote}}^y
& F_{\mathrm{ote}}^z \\
F_{\mathrm{ote}}^z
& F_{\mathrm{ote}}^x + i F_{\mathrm{ote}}^y
\end{array}\right].
\end{align}
Because of the internal spin degree of freedom of the spin-triplet Cooper pairs, we have the three following components:
$F_{\mathrm{ote}}^{\nu}(\boldsymbol{r},\omega)$ for $\nu = x,y,z$.
In Figs.~\ref{fig:figure5}(a) and \ref{fig:figure5}(b), we demonstrate the pair amplitudes at a low-frequency ($\omega=10^{-4}\Delta$)
as a function of the Zeeman potential and superconducting phase difference.
Here, we show the absolute values of (a) $F_{\mathrm{ote}}^{x}(\boldsymbol{r},\omega)$ and (b) $F_{\mathrm{ese}}(\boldsymbol{r},\omega)$,
where the pair amplitudes are averaged in terms of the lattice sites in the DN segment.
As shown in Fig.~\ref{fig:figure5}(a), the OTE Cooper pairs significantly increase in terms of their pair amplitude during the topological phase,
whereas their amplitude during the non-topological phase are almost zero.
Moreover, we confirm that other components of the OTE Cooper pairs (i.e., $F_{\mathrm{ote}}^{y}$ and $F_{\mathrm{ote}}^{z}$)
also have significant amplitudes during the topological phase.
However, as shown in Fig.~\ref{fig:figure5}(b), the pair amplitude of the ESE Cooper pairs is strongly suppressed during the topological phase.
It has been shown that there is an essential duality in Majorana bound states and odd-frequency Cooper pairs~\cite{asano_13,tamura_19}.
Thus, experimental observations of the anomalous proximity effect in the present system will provide remarkable progress regarding
the physics of both Majorana bound states and odd-frequency Cooper pairs.

\textit{Discussion.}
In summary, we demonstrated the anomalous proximity effect of a planer TJJ.
Notably, a TJJ itself has already been realized experimentally~\cite{nichele_19,yacoby_19,shabani_20}.
To experimentally observe the anomalous proximity effect, the thermal coherent length $\xi_T = \sqrt{\hbar D/ 2 \pi k_B T}$
must be longer than the length of the DN segment (i.e., $L_{\mathrm{DN}}$),
where $T$ and $D$ represent the temperature and diffusion constant in the DN segment, respectively.
Nonetheless, because microfabrication techniques for semiconductor thin-films have been well established,
this condition can be satisfied by tuning $L_{\mathrm{DN}}$ or the strength of the potential disorder.
Hence, we propose a highly promising experiment for observing the anomalous proximity effect
related to the essential natures of both Majorana bound states and odd-frequency Cooper pairs.

\begin{acknowledgments}
We are grateful to S. Kashiwaya for the fruitful discussions.
This work was supported by Grants-in-Aid from JSPS for Scientific Research (B) (KAKENHI Grant No. JP18H01176),
and for Scientific Research (A) (KAKENHI Grant No. JP20H00131).
It was also supported by the JSPS Core-to-Core program ``Oxide Superspin" international network.
\end{acknowledgments}

\clearpage
\onecolumngrid
\begin{center}
 \textbf{\large Supplemental Material for \\ ``Anomalous Proximity Effect of Planer Topological Josephson Junctions''}\\ \vspace{0.3cm}
S. Ikegaya$^{1}$, S. Tamura$^{2}$, D. Manske$^{1}$, and Y. Tanaka$^{2}$\\ \vspace{0.1cm}
{\itshape $^{1}$Max-Planck-Institut f\"ur Festk\"orperforschung, Heisenbergstrasse 1, D-70569 Stuttgart, Germany\\
$^{2}$Department of Applied Physics, Nagoya University, Nagoya 464-8603, Japan}
\date{\today}
\end{center}

\section{Bogoliubov-de Gennes Hamiltonian}
\begin{figure}[bbbb]
\begin{center}
\includegraphics[width=0.55\textwidth]{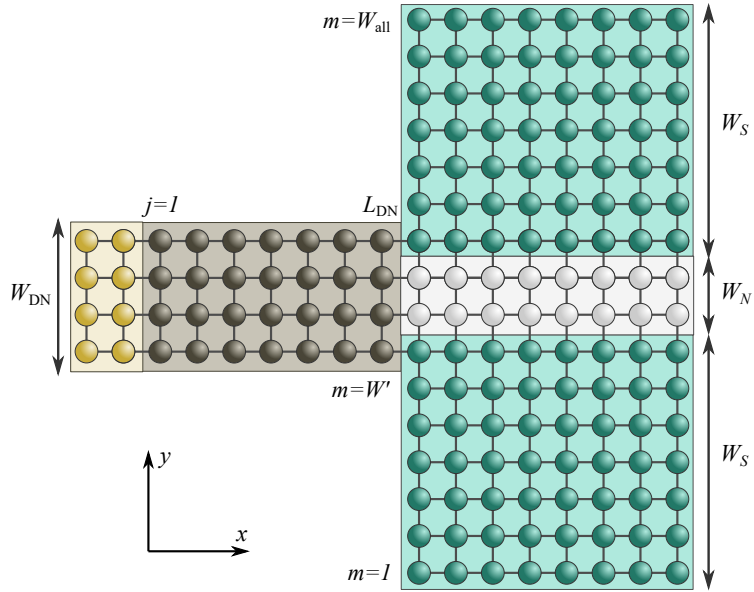}
\caption{Schematic image of the planar topological Josephson junction with the additional dirty normal-metal segment.}
\label{fig:fig_sup}
\end{center}
\end{figure}
In this section, we present the detailed expression for the tight-binding Bogoliubov-de Gennes (BdG) Hamiltonian for
a topological Josephson junction (TJJ) with an additional dirty normal-metal (DN) segment as shown in Fig.~\ref{fig:fig_sup}.
A lattice site is indicated by a vector $\boldsymbol{r}=j \boldsymbol{x} + m \boldsymbol{y}$.
The junction consists of three segment:
a lead wire ($-\infty \leq j \leq 0$), the dirty normal-metal segment ($1 \leq j \leq L_{\mathrm{DN}}$), and the TJJ segment ($L_{\mathrm{DN}}+1 \leq j \leq \infty$).
In the $y$ direction, the TJJ is located between $1 \leq m \leq W_{\mathrm{all}}$,
where $W_{\mathrm{all}}=2W_S + W_N$ with $W_{S(N)}$ representing the width of the superconducting (normal) segment.
The lead wire and DN segment are located for $W^{\prime}+1 \leq m \leq W^{\prime}+W_{\mathrm{DN}}$,
where $W^{\prime} = (W_{\mathrm{all}}-W_{\mathrm{DN}})/2$.
The BdG Hamiltonian reads,
\begin{align}
H_{\mathrm{all}} = H + H_{\mathrm{DN}} + H^{\prime} + H_{\mathrm{int}}.
\label{eq:bdg_ham}
\end{align} 
The first term in Eq.~(\ref{eq:bdg_ham}) describes the TJJ,
\begin{align}
H = & -t \sum_{j=L_{\mathrm{DN}}+1}^{\infty} \; \sum_{m=1}^{W_{\mathrm{all}}} \; \sum_{\sigma = \uparrow, \downarrow}
\left[c^{\dagger}_{\boldsymbol{r}+\boldsymbol{x},\sigma}c_{\boldsymbol{r},\sigma}
+c^{\dagger}_{\boldsymbol{r},\sigma} c_{\boldsymbol{r}+\boldsymbol{x},\sigma} \right]
-t \sum_{j=L_{\mathrm{DN}}+1}^{\infty} \; \sum_{m=1}^{W_{\mathrm{all}}-1} \; \sum_{\sigma}
\left[c^{\dagger}_{\boldsymbol{r}+\boldsymbol{y},\sigma}c_{\boldsymbol{r},\sigma}
+c^{\dagger}_{\boldsymbol{r},\sigma} c_{\boldsymbol{r}+\boldsymbol{y},\sigma} \right] \nonumber\\
&-\mu \sum_{j=L_{\mathrm{DN}}+1}^{\infty} \; \sum_{m=1}^{W_{\mathrm{all}}} \; \sum_{\sigma}
c^{\dagger}_{\boldsymbol{r},\sigma}c_{\boldsymbol{r},\sigma} \nonumber\\
&+\frac{i \lambda}{2} \sum_{j=L_{\mathrm{DN}}+1}^{\infty} \; \sum_{m=1}^{W_{\mathrm{all}}} \; \sum_{\sigma,\sigma^{\prime}}
\left( \sigma_y \right)_{\sigma,\sigma^{\prime}}
\left[c^{\dagger}_{\boldsymbol{r}+\boldsymbol{x},\sigma}c_{\boldsymbol{r},\sigma^{\prime}}
-c^{\dagger}_{\boldsymbol{r},\sigma} c_{\boldsymbol{r}+\boldsymbol{x},\sigma^{\prime}} \right] \nonumber\\
&-\frac{i \lambda}{2}\sum_{j=L_{\mathrm{DN}}+1}^{\infty} \; \sum_{m=1}^{W_{\mathrm{all}}-1} \; \sum_{\sigma,\sigma^{\prime}}
\left( \sigma_x \right)_{\sigma,\sigma^{\prime}}
\left[c^{\dagger}_{\boldsymbol{r}+\boldsymbol{y},\sigma}c_{\boldsymbol{r},\sigma^{\prime}}
-c^{\dagger}_{\boldsymbol{r},\sigma} c_{\boldsymbol{r}+\boldsymbol{y},\sigma^{\prime}} \right] \nonumber\\
&+V_Z\sum_{j=L_{\mathrm{DN}}+1}^{\infty} \; \sum_{m=1}^{W_{\mathrm{all}}} \; \sum_{\sigma, \sigma^{\prime}}
 \left( \sigma_x \right)_{\sigma,\sigma^{\prime}} c^{\dagger}_{\boldsymbol{r},\sigma}c_{\boldsymbol{r},\sigma^{\prime}}, \nonumber\\
&+\sum_{j=L_{\mathrm{DN}}+1}^{\infty} \left[ \sum_{m=1}^{W_S}
\left(\Delta e^{i \varphi/2} c^{\dagger}_{\boldsymbol{r},\uparrow}c^{\dagger}_{\boldsymbol{r},\downarrow} + \mathrm{h.c.}\right)
+\sum_{m=1+W_S}^{W_{\mathrm{all}}}
\left(\Delta e^{-i \varphi/2} c^{\dagger}_{\boldsymbol{r},\uparrow}c^{\dagger}_{\boldsymbol{r},\downarrow} + \mathrm{h.c.}\right) \right],
\label{eq:bdg_ham_tjj}
\end{align}
where $c^{\dagger}_{\boldsymbol{r},\sigma}$ ($c_{\boldsymbol{r},\sigma}$) is the creation (annihilation) operator of an electron
at $\boldsymbol{r}$ with spin $\sigma=\uparrow$, $\downarrow$,
$t$ denotes the nearest-neighbor hopping integral, $\mu$ is the chemical potential.
The strength of the Rashba spin-orbit coupling is represented by $\lambda$.
The Zeeman potential due to the externally applied magnetic field in the $x$ direction is given by $V_Z$.
The amplitude of the pair potential is denoted with $\Delta$, where $\varphi$ represents the superconducting phase difference between the two superconducting segments.
The Pauli matrices in spin space are represented by $\sigma_{\nu}$ for $\nu=x$, $y$, $z$.
The second term in Eq.~(\ref{eq:bdg_ham}) denotes the Hamiltonian for the DN segment,
\begin{align}
H_{\mathrm{DN}} = & -t \sum_{j=1}^{L_{\mathrm{DN}}} \; \sum_{m=W^{\prime}+1}^{W^{\prime}+W_{\mathrm{DN}}} \; \sum_{\sigma}
\left[c^{\dagger}_{\boldsymbol{r}+\boldsymbol{x},\sigma}c_{\boldsymbol{r},\sigma}
+c^{\dagger}_{\boldsymbol{r},\sigma} c_{\boldsymbol{r}+\boldsymbol{x},\sigma} \right]
-t \sum_{j=1}^{L_{\mathrm{DN}}} \; \sum_{m=W^{\prime}+1}^{W^{\prime}+W_{\mathrm{DN}}-1} \; \sum_{\sigma}
\left[c^{\dagger}_{\boldsymbol{r}+\boldsymbol{y},\sigma}c_{\boldsymbol{r},\sigma}
+c^{\dagger}_{\boldsymbol{r},\sigma} c_{\boldsymbol{r}+\boldsymbol{y},\sigma} \right] \nonumber\\
&-\mu\sum_{j=1}^{L_{\mathrm{DN}}} \; \sum_{m=W^{\prime}+1}^{W^{\prime}+W_{\mathrm{DN}}} \; \sum_{\sigma}
c^{\dagger}_{\boldsymbol{r},\sigma}c_{\boldsymbol{r},\sigma} \nonumber\\
&+\frac{i \lambda}{2}\sum_{j=1}^{L_{\mathrm{DN}}} \; \sum_{m=W^{\prime}+1}^{W^{\prime}+W_{\mathrm{DN}}} \; \sum_{\sigma,\sigma^{\prime}}
\left( \sigma_y \right)_{\sigma,\sigma^{\prime}}
\left[c^{\dagger}_{\boldsymbol{r}+\boldsymbol{x},\sigma}c_{\boldsymbol{r},\sigma^{\prime}}
-c^{\dagger}_{\boldsymbol{r},\sigma} c_{\boldsymbol{r}+\boldsymbol{x},\sigma^{\prime}} \right] \nonumber\\
&-\frac{i \lambda}{2}\sum_{j=1}^{L_{\mathrm{DN}}} \; \sum_{m=W^{\prime}+1}^{W^{\prime}+W_{\mathrm{DN}}-1} \; \sum_{\sigma, \sigma^{\prime}}
\left( \sigma_x \right)_{\sigma,\sigma^{\prime}}
\left[c^{\dagger}_{\boldsymbol{r}+\boldsymbol{y},\sigma}c_{\boldsymbol{r},\sigma^{\prime}}
-c^{\dagger}_{\boldsymbol{r},\sigma} c_{\boldsymbol{r}+\boldsymbol{y},\sigma^{\prime}} \right] \nonumber\\
&+V_Z\sum_{j=1}^{L_{\mathrm{DN}}} \; \sum_{m=W^{\prime}+1}^{W^{\prime}+W_{\mathrm{DN}}} \; \sum_{\sigma, \prime}
 \left( \sigma_x \right)_{\sigma,\sigma^{\prime}} c^{\dagger}_{\boldsymbol{r},\sigma}c_{\boldsymbol{r},\sigma^{\prime}} \nonumber\\
&+\mu\sum_{j=1}^{L_{\mathrm{DN}}} \; \sum_{m=W^{\prime}+1}^{W^{\prime}+W_{\mathrm{DN}}} \; \sum_{\sigma}
v(\boldsymbol{r}) c^{\dagger}_{\boldsymbol{r},\sigma}c_{\boldsymbol{r},\sigma},
\end{align}
where $v(\boldsymbol{r})$ is the disordered potential given randomly in the range of $-X \leq v(\boldsymbol{r}) \leq X$.
The third term of Eq.~(\ref{eq:bdg_ham}) describes the lead wire,
\begin{align}
H^{\prime} = & -t^{\prime} \sum_{j=-\infty}^{-1} \; \sum_{m=W^{\prime}+1}^{W^{\prime}+W_{\mathrm{DN}}} \; \sum_{\sigma}
\left[c^{\dagger}_{\boldsymbol{r}+\boldsymbol{x},\sigma}c_{\boldsymbol{r},\sigma}
+c^{\dagger}_{\boldsymbol{r},\sigma} c_{\boldsymbol{r}+\boldsymbol{x},\sigma} \right]
-t^{\prime} \sum_{j=-\infty}^{-1} \; \sum_{m=W^{\prime}+1}^{W^{\prime}+W_{\mathrm{DN}}-1} \; \sum_{\sigma}
\left[c^{\dagger}_{\boldsymbol{r}+\boldsymbol{y},\sigma}c_{\boldsymbol{r},\sigma}
+c^{\dagger}_{\boldsymbol{r},\sigma} c_{\boldsymbol{r}+\boldsymbol{y},\sigma} \right] \nonumber\\
&-\mu^{\prime} \sum_{j=-\infty}^{-1} \; \sum_{m=W^{\prime}+1}^{W^{\prime}+W_{\mathrm{DN}}} \; \sum_{\sigma}
c^{\dagger}_{\boldsymbol{r},\sigma}c_{\boldsymbol{r},\sigma} \nonumber\\
&+V_Z^{\prime} \sum_{j=-\infty}^{-1} \; \sum_{m=W^{\prime}+1}^{W^{\prime}+W_{\mathrm{DN}}} \; \sum_{\sigma, \prime}
 \left( \sigma_x \right)_{\sigma,\sigma^{\prime}} c^{\dagger}_{\boldsymbol{r},\sigma}c_{\boldsymbol{r},\sigma^{\prime}},
\end{align}
where $t^{\prime}$, $\mu^{\prime}$ and $V^{\prime}_Z$ represent the hopping integral, chemical potential, and Zeeman potential in the normal wire, respectively.
The last term of  Eq.~(\ref{eq:bdg_ham}) gives the coupling between the lead wire and the DN segment,
\begin{align}
H_{\mathrm{int}} = -t_{\mathrm{int}} \sum_{m=W^{\prime}+1}^{W^{\prime}+W_{\mathrm{DN}}} \; \sum_{\sigma}
\left[c^{\dagger}_{j=1,m,\sigma}c_{j=0,m,\sigma} +c^{\dagger}_{j=0,m,\sigma} c_{j=1,m,\sigma} \right],
\end{align}
where the hopping integral between the lead wire and the DN segment is given by $t_{\mathrm{int}}$.

\section{Symmetry of Topological Josephson Junction}
In this section, we discuss the symmetry properties of the TJJ.
We here remove the DN segment and lead wire from the TJJ.
By applying the periodic boundary condition in the $x$ direction,
the BdG Hamiltonian of the TJJ, which is given in Eq.~(\ref{eq:bdg_ham_tjj}), can be deformed as
\begin{gather}
H = \frac{1}{2} \sum_{k_x} \boldsymbol{C}^{\dagger}_{k_x} \; \check{H}(k_x) \; \boldsymbol{C}^{\dagger}_{k_x},\nonumber\\
\boldsymbol{C}_{k_x}=\left[\boldsymbol{C}_{k_x\uparrow},\boldsymbol{C}_{k_x\downarrow},
\boldsymbol{C}^{\dagger}_{k_x\uparrow},\boldsymbol{C}^{\dagger}_{k_x\downarrow}\right]^{\mathrm{T}}, \qquad
\boldsymbol{C}_{k_x\sigma} = \left[c_{k_x,1,\sigma},c_{k_x,2,\sigma}, \cdots c_{k_x,W_{\mathrm{all}},\sigma} \right]^{\mathrm{T}},\\
\check{H}(k_x) = \left[ \begin{array}{cc}
\hat{h}(k_x) & \hat{\Delta} \\
-\hat{\Delta}^{\ast} & -\hat{h}^{\ast}(-k_x) \\
\end{array} \right],\nonumber
\end{gather}
with
\begin{align}
\begin{split}
&\hat{h}(k_x) = \left[ \begin{array}{cc}
\bar{\xi}(k_x) & - i \bar{\lambda}_x(k_x) - \bar{\lambda}_y + \bar{V}_Z \\
i \bar{\lambda}_x(k_x) - \bar{\lambda}_y + \bar{V}_Z & \bar{\xi}(k_x) \\
\end{array} \right],\\
&\bar{\xi}(k_x) = \left\{ 2t \left(1-\cos k_x \right) + 2t - \mu \right\} \bar{I} +
\left[ \begin{array}{ccccc}
0 & -t& &  & \mbox{\Large 0} \\
-t & 0 & -t & & \\
  & \ddots& \ddots & \ddots & \\
  & & -t &0 & -t \\
 \mbox{\Large 0} &  &  & -t & 0\\
\end{array}\right], \quad
\bar{\lambda}_y = 
\left[ \begin{array}{ccccc}
0 & -i \lambda/2& &  & \mbox{\Large 0} \\
i \lambda/2 & 0 & -i \lambda/2 & & \\
  & \ddots& \ddots & \ddots & \\
  & & i \lambda/2 &0 & -i \lambda/2 \\
 \mbox{\Large 0} &  &  & i \lambda/2 & 0\\
\end{array}\right],\\
&\bar{\lambda}_x(k_x) = \lambda \sin k_x \bar{I}, \qquad
\bar{V}_Z = V_Z \bar{I},
\end{split}
\end{align}
and
\begin{align}
\hat{\Delta} = \left[ \begin{array}{cc}
0 & \bar{\Delta} \\
-\bar{\Delta} & 0 \\
\end{array} \right], \quad
\bar{\Delta} = \left[ \begin{array}{ccc}
\tilde{\Delta}_+ &  &  \mbox{\Large 0} \\
& O_N & \\
 \mbox{\Large 0} & & \tilde{\Delta}_-
\end{array} \right], \quad
\tilde{\Delta}_{\pm} = \left[ \begin{array}{ccc}
\Delta e^{\pm i\varphi/2} &  &  \mbox{\Large 0} \\
& \ddots & \\
 \mbox{\Large 0} & & \Delta e^{\pm i \varphi/2}
\end{array} \right],
\end{align}
where $c^{\dagger}_{k_x,m,\sigma}$ ($c_{k_x,m,\sigma}$) is the creation (annihilation) operator of an electron at $y = m a_0$ layer with momentum $k_x$ and spin $\sigma$.
$\bar{A}$ ($\tilde{A}$) represents a $W_{\mathrm{all}} \times W_{\mathrm{all}}$ ($W_S \times W_S$) matrix.
$\bar{I}$ and $O_N$ represent the $W_{\mathrm{all}} \times W_{\mathrm{all}}$ unit matrix and $W_N \times W_N$ zero matrix, respectively.

The standard time-reversal symmetry operator in this basis is given by
\begin{align}
\check{T}_- = \left[ \begin{array}{cc}
i \hat{\sigma}_y & 0 \\ 0 & i \hat{\sigma}_y \end{array} \right] {\cal K}, \qquad
\hat{\sigma}_y = \left[ \begin{array}{cc}
0 &  - i \bar{I} \\ i \bar{I} & 0 \end{array} \right],
\end{align}
with obeying $\check{T}_-=-1$, where ${\cal K}$ represents the complex-conjugation operator.
The mirror reflection symmetry operator with respect to the $x$-$z$ plane is defined as
\begin{align}
\check{M}_y = \left[ \begin{array}{cc}
\hat{M}_y & 0 \\ 0 & \hat{M}^{\ast}_y \end{array} \right] , \qquad
\hat{M}_y = i  \left[ \begin{array}{cc}
0 &  - i \bar{P}_y \\  i \bar{P}_y & 0 \end{array} \right], \qquad
\bar{P}_y = \left[ \begin{array}{ccc}
\mbox{\large 0} & & 1\\
& \iddots & \\
1 & & \mbox{\large 0} \end{array} \right],
\end{align}
with obeying $\check{M}^2_y=-1$, where the mirror reflection in the spatial coordinate is described by $\bar{P}_y$.
The BdG Hamiltonian $\check{H}(k_x)$ does not have the standard time-reversal symmetry of $\check{T}_-$
and mirror reflection symmetry of $\check{M}_y$.
Nevertheless, $\check{H}(k_x)$ satisfies
\begin{align}
\begin{split}
&\check{T}_+ \; \check{H}(k_x) \; \check{T}^{-1}_+ = \check{H}(-k_x), \\
&\check{T}_+ = \check{M}_y \check{T}_+ = - \left[ \begin{array}{cc}
\hat{P}_y & 0 \\ 0 & \hat{P}_y \end{array} \right] {\cal K}, \quad
\hat{P}_y = \left[ \begin{array}{cc}
\bar{P}_y & 0 \\ 0 & \bar{P}_y \end{array} \right],\\
\end{split}
\end{align}
which represents time-reversal symmetry of $\check{H}(k_x)$ with $\check{T}^2_+=+1$.
The BdG Hamiltonian intrinsically preserves particle-hole symmetry as
\begin{align}
\check{C} \; \check{H}(k_x) \; \check{C}^{-1}_+ = - \check{H}(-k_x), \quad
\check{C}_+  = - \left[ \begin{array}{cc}
0 & \hat{I} \\ \hat{I} & 0 \end{array} \right] {\cal K},
\end{align}
with obeying $\check{C}^2=+1$, where $\hat{I}$ represent $2 W_{\mathrm{all}} \times 2 W_{\mathrm{all}}$ unit matrix.
By combining $\check{T}_+$ and $\check{C}$, we obtain chiral symmetry of $\check{H}(k_x)$ as
\begin{align}
\check{S} \; \check{H}(k_x) \; \check{S}^{-1} = -\check{H}(k_x), \qquad
&\check{S} = -\check{T}_+ \check{C} = - \left[ \begin{array}{cc}
0 & \hat{P}_y \\ \hat{P}_y  & 0 \end{array} \right].
\end{align}
Since $\check{T}^2_+ = +1$ and $\check{C}^2 = +1$, the BdG Hamiltonian $\check{H}(k_x)$ belongs to the BDI symmetry class.
Thus, we can defined the one-dimensional winding number calculated by
\begin{align}
w = \frac{1}{4 \pi i} \int dk_x \mathrm{Tr} \left[ \check{S} \check{H}^{-1}(k_x) \partial_{k_x} \check{H}(k_x) \right],
\end{align}
which is also given in the main text.

\end{document}